\documentclass[sigconf,anonymous=false]{acmart}

\renewcommand\footnotetextcopyrightpermission[1]{}
\usepackage{booktabs} 
\usepackage{flexisym}
\usepackage{bbm}
\usepackage{makecell}

\usepackage{balance}
\usepackage{graphicx}
\usepackage{multirow}
\usepackage{float}

\usepackage{figsize}
\usepackage{amsmath}
\usepackage{todonotes}

\newcommand{\norm}[1]{\left\lVert#1\right\rVert}

\copyrightyear{2020}
\acmYear{2020}
\setcopyright{acmcopyright}
\acmConference{WSDM'20}{}{February 3--7, 2020, Houston, Texas, USA} 
\acmPrice{15.00} 
\acmDOI{http******}
\acmISBN{ISBN *************}

\renewcommand\footnotetextcopyrightpermission[1]{}

\begin{document}
\fancyhead{}

\title{The Impact of Popularity Bias on Fairness and Calibration in Recommendation}

\author{Himan Abdollahpouri}
\affiliation{%
  \institution{University of Colorado Boulder}
  \country{USA}
}
\email{himan.abdollahpouri@colorado.edu}
\author{Masoud Mansoury}
\affiliation{%
  \institution{Eindhoven University of Technology}
  \country{Eindhoven, the Netherlands} 
}
\email{m.mansoury@tue.nl}
\author{Robin Burke}
\affiliation{%
  \institution{University of Colorado Boulder}
  \country{USA}
}
\email{robin.burke@colorado.edu}
\author{Bamshad Mobasher}
\affiliation{%
  \institution{DePaul University}
  \country{USA}
}
\email{mobasher@cs.depaul.edu}
\begin{abstract}

Recently there has been a growing interest in fairness-aware recommender systems, including fairness in providing consistent performance across different users or groups of users. A recommender system could be considered unfair if the recommendations do not fairly represent the tastes of a certain group of users while other groups receive recommendations that are consistent with their preferences. In this paper, we use a metric called miscalibration for measuring how a recommendation algorithm is responsive to users' true preferences and we consider how various algorithms may result in different degrees of miscalibration. A well-known type of bias in recommendation is popularity bias where few popular items are over-represented in recommendations, while the majority of other items do not get significant exposure. We conjecture that popularity bias is one important factor leading to miscalibration in recommendation. Our experimental results using two real-world datasets show that there is a strong correlation between how different user groups are affected by algorithmic popularity bias and their level of interest in popular items. Moreover, we show that the more a group is affected by the algorithmic popularity bias, the more their recommendations are miscalibrated. Finally, we show that the algorithms with greater popularity bias amplification tend to have greater overall miscalibration.  
\end{abstract}


%
%
\maketitle

\section{Introduction}

Recommender systems have been widely used in a variety of different domains such as e-commerce, movies, music, entertainment, and online dating. Their goal is to help users find relevant items which are difficult or otherwise time-consuming to find in the absence of such systems.
Recommendations are typically evaluated using measures such as precision, diversity, and novelty. Under such measures, depending on the situation, a recommended list of items may be considered good if it is relevant to the user, is diverse and also helps the user discover products that s/he would have not been able to discover in the absence of the recommender system. 

One of the important considerations in recommendation generation that has emerged recently is fairness. Recommendation fairness may have different meanings depending on the domain in which the recommender system is operating, the characteristics of different users or groups of users (e.g. protected vs unprotected), and the goals of the system designers. For instance, Ekstrand et al. \cite{ekstrand2018all} defined fairness as consistent accuracy across different groups of users. In their experiments, they observed certain groups such as women get lower accuracy results than men.

One of the metrics used to measure recommendation quality is calibration, which measures whether the recommendations delivered to a user are consistent with the spectrum of items the user has previously rated. For example, if a user has rated 70\% action movies and 30\% romance, the user might expect to see a similar pattern in the recommendations \cite{steck2018calibrated}. If this ratio differs from the one in the user's profile, we say the recommendations are miscalibrated. Miscalibration by itself may not be considered unfair as it could be simply mean the recommendations are not personalized enough. However, if different users or groups of users experience different levels of miscalibration in their recommendations, this may indicate an unfair treatment of a group of users. For example, authors in \cite{DBLP:journals/corr/YaoH17} define several fairness metrics which focus on having a consistent performance in terms of estimation error across different user groups.

One well-known limitation of collaborative recommender systems is the problem of popularity bias \cite{longtailnichesriche}: popular items are recommended frequently, in some cases even more than their popularity would warrant, while the majority of other items do not get proportional attention. We define algorithmic popularity bias as the tendency of an algorithm to amplify existing popularity differences across items in this way. We measure this amplification through the metric of \textit{popularity lift}, which quantifies the difference between average item popularity in input and output for an algorithm.
Such popularity bias could be problematic for a variety of different reasons: long-tail (non-popular) items are important for generating a fuller understanding of users' preferences ~\cite{nguyen2014exploring, resnick2013bursting}. In addition, long-tail recommendation can also be understood as a social good; a market that suffers from popularity bias will lack opportunities to discover more obscure products and will be, by definition, dominated by a few large brands or well-known artists~\cite{celma2008hits}. Such a market will be more homogeneous and offer fewer opportunities for innovation and creativity.

In this paper, we conjecture that popularity bias is an important factor leading to miscalibration of the recommendation lists. We also show that users with different level of interest in popular items get different level of miscalibration and hence resulting in an unfair treatment by the recommendation algorithms for different groups of users. We show that, across multiple algorithms, there is a strong negative association between users' interest in popular items and the popularity lift of their recommendation. Users with less interest in popular items are the most affected by popularity bias. In addition, we show that algorithms with higher popularity bias tend to have also higher miscalibration, showing the correlation between popularity bias and miscalibration in recommendation algorithms.

Our contributions are as follows:
\begin{itemize}

 \item \textbf{Differential impact of popularity bias:} We show that different groups of users are affected differently by popularity bias.
 \item \textbf{Connection between popularity bias and miscalibration:} We show that algorithms with a higher popularity amplification tend to also a have higher degree of overall miscalibration.
 \item \textbf{Association of popularity bias with miscalibration:} We show that when the popularity lift is higher for a group, its miscalibration is also higher. 
\end{itemize}

\section{Related Work}
The problem of popularity bias and the challenges it creates for the recommender system has been well studied by other researchers \cite{anderson2006long,brynjolfsson2006niches,longtailrecsys}. Authors in the mentioned works have mainly explored the overall accuracy of the recommendations in the presence of long-tail distribution in rating data. Moreover, some other researchers have proposed algorithms that can control this bias and give more chance to long-tail items to be recommended \cite{10.1109/TKDE.2011.15,DBLP:conf/recsys/KamishimaAAS14, abdollahpouri2017controlling,flairs2019}. 

Moreover, the concept of fairness in recommendation has been also gaining a lot of attention recently \cite{kamishima2012fairness,DBLP:journals/corr/YaoH17}. For example, finding solutions that remove algorithmic discrimination against users belong to a certain demographic information \cite{zhu2018fairness} or making sure items from different categories (e.g. long tail items or items belong to different providers)\cite{liu2018personalizing,burke2017balanced} are getting a fair exposure in the recommendations. Our definition of fairness in this paper is aligned with the fairness objectives introduced by Yao and Huang in \cite{DBLP:journals/corr/YaoH17} where they define unfairness as having inconsistent estimation error across different users. We can generalize the \textit{estimation error} to simply be any kind of system performance. For instance, Steck defines fairness as concerning the various interests of a user, with the goal to reflecting them
according to their corresponding proportions \cite{steck2018calibrated} (i.e. calibration). In this paper, we use the same definition for fairness: a system is \textit{unfair} if it delivers different degree of miscalibration to different users. 

With regard to looking at the performance of the recommender system for different user groups, Ekstrand et al. \cite{ekstrand2018all} showed that some recommendation algorithms give significantly less accurate recommendations to groups from certain age or gender. In addition, Abdollahpouri et al. in \cite{abdollahpouri2019beyond} discuss the importance of recommendation evaluation with respect to the distribution of utilities given to different stakeholders. For instance, the degree of calibration of the recommendation for each user group (i.e. a stakeholder) could be considered as its utility and therefore, a balanced distribution of these utility values is important in a fair recommender system. 

And finally, Jannach et al. \cite{jannach2015recommenders} compared different recommendation algorithms in terms of accuracy and popularity bias. In that paper they observed some algorithms concentrate more on popular items than the others. In our work, we are mainly interested in seeing the popularity bias from the users' expectations perspective. Our work is the first attempt in connecting popularity bias and the concept of miscalibration. 
\section{Popularity Bias and Miscalibration}
Popularity bias and miscalibration are both aspects of an algorithm's performance that are computed by comparing the attributes of the input data with the properties of the recommendations that are produced for users. In this section, we define these terms more precisely.

\subsection{Miscalibration}
One of the interpretations of fairness in recommendation is in terms of whether the recommender provides consistent performance across different users or groups of users. A recommender system could be considered unfair if the recommendations do not fairly represent the tastes of a certain group of users while other groups receive recommendations that are consistent with their preferences. In this paper we use a metric called miscalibration \cite{steck2018calibrated} for measuring how a recommendation algorithm is responsive to users' true preferences and we consider how various algorithms may result in different degrees of miscalibration. As we mentioned earlier, miscalibration, if it exists across all users, could simply mean failure of the algorithm to provide accurate  personalization. But when different groups of users experience different levels of miscalibration, this could indicate unfair treatment of certain user groups.  From this standpoint, we call a recommender system unfair if it has different levels of miscalibration for different user groups.

\begin{figure}
\centering
\SetFigLayout{1}{2}
  \subfigure[The inconsistency of recommender system's performance for users A and B in terms of miscalibration. User B's recommendations are highly consistent (calibrated) with her profile while user A's interest has been distorted in the recommendations.]{\includegraphics[width=3.3in]{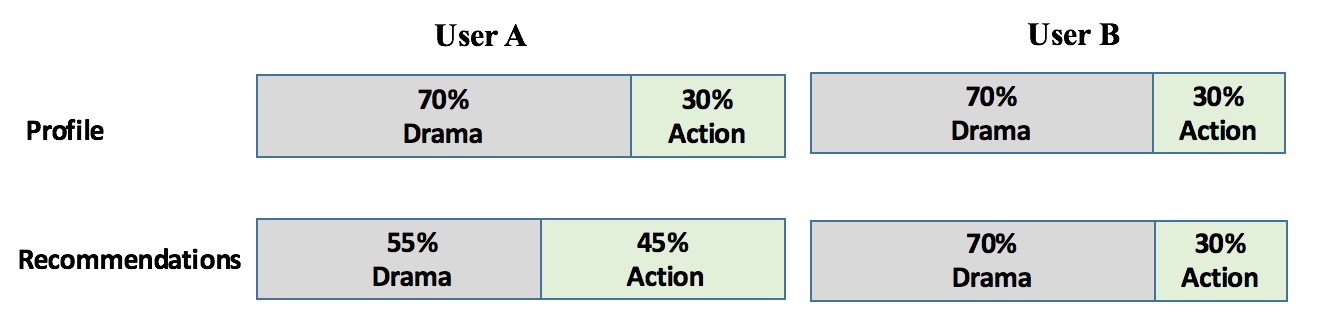}}
  \hfill
  \subfigure[The inconsistency of recommender system's performance for users A and B in terms of popularity bias. For user B there is no amplification of popularity bias while user A has been highly affected.]{\includegraphics[width=3.3in]{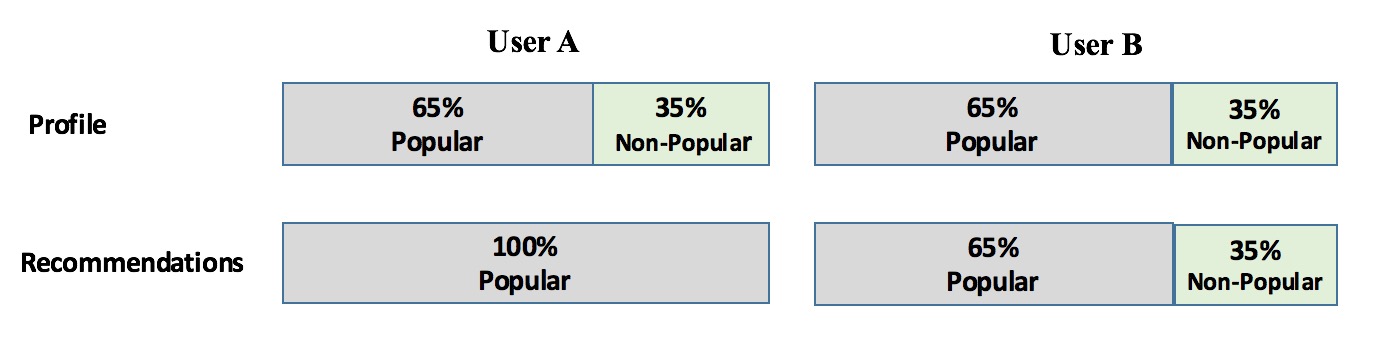}}
  \hfill
\caption{Two different types of inconsistency of recommender system's performance across different users}
\label{calibration}
\end{figure}

Calibration is a general concept in machine learning, and recently
experienced a resurgence in the context of fairness of machine
learning algorithms. A classification algorithm is called calibrated if the predicted proportions of the various classes agree with the actual proportions of classes in the training data. Extending this notion to recommender systems, a calibrated recommender system is one that reflects the various interests of a user in the recommended list, and with
their appropriate proportions. Figure ~\ref{calibration}a shows two users, A and B, and their profiles. User A has 70\% Drama movies and 30\% Action movies in her profile. Therefore, it is expected to see the same ratio in her recommendations but, as you can see, this proportion has been distorted by giving 55\% (less than it was expected) drama movies and 45\% (more than it was expected) action movies as recommendations to that user. On the other hand, user B has also rated 70\% Drama and 30\% Action and her recommendations perfectly match her expectations. This shows an unfair scenario where the recommender system does not perform as well for user A as it does for user B. 

For measuring the miscalibration of the recommendations we use the metric introduced in \cite{steck2018calibrated}.Assume $u$ be a user and $i$ be an item. Also, suppose for each item $i$ there is a set of features $C$ describing the item. For example, a song could be pop, jazz or a movie could have genres action, romance, comedy, etc. We use $c$ for each of these individual categories. Also, we assume that each user has rated one or more items, showing interest in features $c$ belonging to those items. We consider two distributions for each user $u$, one corresponding to the distribution of categories $c$ across all items rated by $u$, and another representing the distribution of categories $c$ across all recommended items to $u$:

\begin{itemize}
    \item $p_u(c|u)$: the distribution over feature $c$ of the set of items $\Gamma$ rated by user $u$ in the past:
    
\begin{equation}
    p_u(c|u)=\frac{\sum_{i \in \Gamma}w_{u,i}. p(c|i)}{\sum_{i \in \Gamma}w_{u,i}}
\end{equation}

where $w_{u,i}$ is the weight of item $i$, e.g., how recently it was rated by user $u$. In this paper, we set $w$ to 1 to focus more specifically on differences in feature distribution rather than temporal aspects of user profiles. 

    \item $q_u(c|u)$: the distribution over feature $c$ of the list of items
recommended to user $u$: 

\begin{equation}
    q_u(c|u)=\frac{\sum_{i \in \Lambda}w_r(i). p(c|i)}{\sum_{i \in \Lambda}w_r(i)}
\end{equation}

where $\Lambda$ is the set of recommended items. The weight
of item $i$ due to its rank $r(i)$ in the recommendations is
denoted by $w_r(i)$. Possible choices include the weighting
schemes used in ranking metrics, like in Mean Reciprocal
Rank (MRR) or normalized Discounted Cumulative Gain
(nDCG). As above, we set the weight $w_r$ to 1 in order to focus on feature distribution across the recommendation set and to ensure the $q_u$ values are comparable to the $p_u$ values.

\end{itemize}

The degree of dissimilarity between  $q_u(c|u)$ and $p_u(c|u)$ is used to compute miscalibration in recommendations.  
There are various established methods for determining if two
finite distributions are similar, such as statistical hypothesis testing, with
the null hypothesis being that the two distributions are the same, as mentioned in \cite{steck2018calibrated}. Authors in that paper used Kullback-Leibler (KL) divergence as miscalibration metric. In our data, there are many profiles with no ratings for some subset of the features, leading to zero values in the $p_u$ distribution, and similarly, recommendation lists may concentrate only on certain features, causing zero values in the $q_u$ distribution for some users. $KL$ divergence is undefined where there are no observations. As an alternative, we use the Hellinger distance, $H$, as suggested by authors in \cite{steck2018calibrated} for situations where we have many zeros. So, we measure miscalibration for user $u$, $MC_(p,q)_u$  as follows:

\begin{equation}
    MC_u(p_u,q_u)=H(p_u,q_u)= \frac{\norm{\sqrt{p_u}-\sqrt{q_u}}_2}{\sqrt{2}}
\end{equation}
By definition, the Hellinger distance is a metric satisfying triangle inequality. The $\sqrt{2}$
 in
the definition is for ensuring that $H(p_u, q_u) \leq 1$ for all probability distributions \cite{cieslak2012hellinger}.

The overall miscalibration metric $MC_G$ for each group $G$ is obtained by averaging $MC_u(p,q)$
across all users $u$ in group $G$. That is:

\begin{equation}
    MC_G(p,q)=\frac{\sum_{u \in G} MC_u(p_u,q_u)}{|G|}
\end{equation}

$MC_G(p,q)$ is 0 when $p$ and $q$ are similar and it increases when their difference is higher.  
\subsubsection{Fairness}
Similar to \cite{steck2018calibrated}, in this paper we define a system to be \textit{unfair} if it gives different levels of miscalibration to different user groups. One could define a group of users in many different ways based on some features such as gender, age, occupation, education etc. Likewise, we could define a group of users based on how similar their interests are. For instance, authors in \cite{abdollahpourirmse1} categorize users into different groups based on their degree of interest in popular items (i.e. if users are interested in niche movies versus blockbuster ones). In this paper we use the same grouping methodology because we are interested in measuring how popularity bias impacts users with different level of interest in popular items and also how it is correlated with the miscalibration of the recommendation lists for different users. Some insights based on grouping using gender information is also reported in Section 6.  
\subsection{Popularity Bias in Rating Data}

As noted earlier, recommendation algorithms are known to suffer from popularity bias. That is, a few items are recommended to many users while the majority of other items do not get a deserved exposure. This bias could be because of the inherent nature of rating data which is skewed towards popular items and also because of the algorithmic amplification of such bias. Figure ~\ref{calibration}-b shows the percentage of rated items by two users $A$ and $B$. We can see that user A's recommendations are highly affected by popularity bias while for user B there is no amplification of popularity bias in her recommendations. 

\begin{figure}
\centering
\SetFigLayout{1}{2}
  \subfigure[MovieLens]{\includegraphics[width=1.653in]{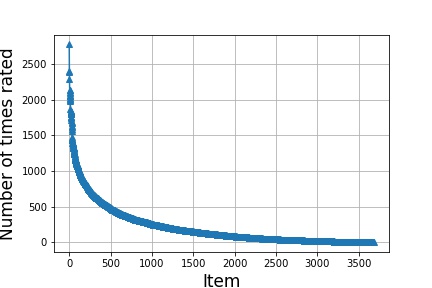}}
  \hfill
  \subfigure[Yahoo Movies]{\includegraphics[width=1.653in]{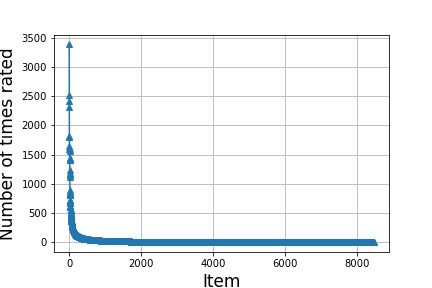}}
  \hfill
\caption{The long-tail of item popularity in rating data in MovieLens and Yahoo Movies datasets.}
\label{longtail}
\end{figure}

In many domains, rating data is skewed towards more popular items--there are a few popular items with the majority of ratings while the rest of the items have far fewer ratings. Figure ~\ref{longtail} shows the long-tail distribution of item popularity in the well known MovieLens 1M  and Yahoo movies datasets. Similar distributions can be found in other datasets as well. Although it is true that popular items are popular for a reason, algorithmic popularity bias often amplifies this bias to a great extent.

\begin{figure}
\centering
\SetFigLayout{1}{2}
  \subfigure[MovieLens]{\includegraphics[width=1.6in]{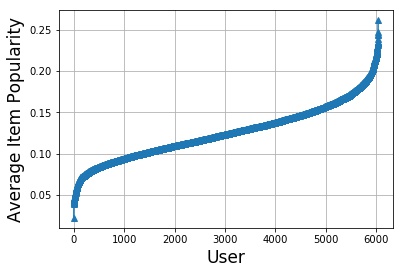}}
  \hfill
  \subfigure[Yahoo Movies]{\includegraphics[width=1.6in]{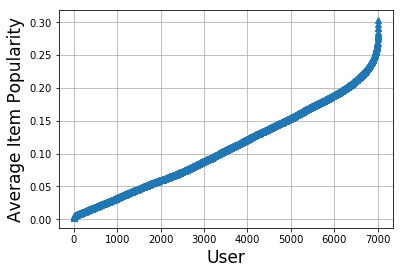}}
  \hfill
\caption{Users propensity towards popular items in MovieLens and Yahoo Movies datasets.}
\label{user_propensity}
\end{figure}

Not every user has the same degree of interest towards popular items \cite{oh2011novel,abdollahpourirmse1}. There might be users who are interested in less popular, niche items. The recommender algorithm should be able to address the needs of those users as well. Figure ~\ref{user_propensity} shows the average popularity of rated items in different users' profiles for both MovieLens and Yahoo Movie datasets. The users have been sorted first based on the average popularity of items in their profile then the data has been plotted. On Movielens dataset there are few users with extreme average item popularity on the most right-hand and left-hand side of the plot while the majority of the users fall in the middle of the distribution with average item popularity between 0.10 and 0.15. In Yahoo Movies, there are more users with low-popularity profiles, but otherwise, the distribution is similar. Both of these plots confirm that users have different degrees of interest towards popular items. 

Due to this common imbalance in the original rating data, often algorithms propagate and, in many cases, amplify the bias by over-recommending the popular items, giving them a higher opportunity for being rated by more users. Repeated recommendation cycles will continue to perpetuate this bias resulting in a  rich-get-richer and poor-get-poorer vicious cycle. However, not every recommendation algorithm has the same amplification power for popularity bias. In the next sections, we will define a metric for measuring the degree to which popularity bias is propagated by the recommendation algorithm. We will empirically evaluate the performance of different algorithms with respect to popularity bias amplification. We will also empirically show the relationship between popularity bias and miscalibration across different algorithms and data sets.

\section{Methodology}

We conducted our experiments on two publicly available datasets. The first one is MovieLens 1M dataset which contains 1,000,209 anonymous ratings of approximately 3,900 movies made by 6,040 MovieLens users~\cite{movielens}. Each movie is associated with at least one genre in this dataset with a total of 18 unique genres in the entire dataset. The second dataset we used is Yahoo Movies\footnote{https://webscope.sandbox.yahoo.com/catalog.php?datatype=r} which originally contained 211,231 ratings provided by 7,642 users on 11,916 movies. There are many items with only one rating making this dataset extremely sparse. Therefore, we created a core-10 sample (each user and each movie should have at least 10 ratings). Our final sample of Yahoo Movies dataset contains 173,676 ratings on 2,131 movies provided by 7,012 users. Analogous to MovieLens dataset, in Yahoo Movies dataset, each movie is associated with at least one genre with a total of 24 genres in the entire datatset. For all experiments, we set aside a random selection of 80\% of the rating data as training set and the remaining 20\% as the test set.  
The reason we use these two datasets is that they both have meaningful features associated with items that could represent users' taste which is needed for measuring the miscalibration. For example, the genre of a movie is indeed a meaningful characteristic of a movie which could represent users' taste in movies. 

We used several recommendation algorithms from the open source java recommendation library (Librec) \cite{guo2015librec} including user-based collaborative filtering ($UserKNN$) \cite{resnick1997recommender}, item-based collaborative filtering ($ItemKNN$) \cite{sarwar2001item}, singular value decomposition (\textit{SVD++}) \cite{koren2008factorization}, and biased matrix factorization ($BMF$) \cite{koren2009matrix} to cover both neighborhood based and matrix factorization based algorithms. We also included the most-popular method (a non-personalized algorithm recommending the most popular items to every user) as an algorithm with extreme popularity bias. We tuned each algorithm to achieve its best performance in terms of precision. Table ~\ref{precision} shows the precision values for each algorithm on both MovieLens and Yahoo Movies datasets. On MovieLens data, $ItemKNN$ has the highest precision while on Yahoo Movies, \textit{SVD++} outperforms other algorithms. We set the size of the generated recommendation list for each user to 10.  

\begin{table}
\caption{$Precision@10$ for different algorithms on MovieLens and Yahoo Movies datasets.}
\begin{tabular}{l*{3}{c}r}
\bottomrule
 algorithm & MovieLens & Yahoo Movies \\
\hline
ItemKNN & 0.223 & 0.127 \\
UserKNN            & 0.214 & 0.13  \\
Most popular           & 0.182 & 0.1 \\
SVD++     & 0.122 & 0.2 \\
BMF     & 0.107 & 0.047 \\
\bottomrule
\end{tabular}
\label{precision}
\end{table}

In this paper, we are interested in seeing how different groups of users with varying degree of interest towards popular items are treated by the recommender system. Therefore, we grouped users in both datasets into an arbitrary number of groups (10 in this paper) based on their degree of interest in popular items. That is, we first measure the average popularity of the rated items in each user's profile and then organize them into 10 groups with the first group having the lowest average item popularity (extremely niche users) and the last group with the highest average item popularity (heavily blockbuster-focused users). We denoted these groups as $G_1$ through $G_{10}$. In Section 6 we also discuss grouping based on gender. 

In order to measure how each algorithm amplifies the popularity bias in its generated recommendations for different user groups, we define \textit{popularity lift} as a measure of the amplification power for different algorithms on different user groups. First, we measure the average item popularity of a group G (i.e. Group Average Popularity) as follows:

\begin{equation}
    GAP_p(G)=\frac{\sum_{u \in G} \frac{\sum_{i \in \Gamma_u}\theta(i)}{|\Gamma_u|}  }{|G|}
\end{equation}

where $\theta(i)$ is the popularity value for item $i$ (i.e. the ratio of users who rated that item) and subscript $p$ refers to the profile of users. 

We also measure the average popularity of the recommended items to the users in the same group as:

\begin{equation}
    GAP_q(G)=\frac{\sum_{u \in G} \frac{\sum_{i \in \Lambda}\theta(i)}{|\Lambda_u|}  }{|G|}
\end{equation}
where subscript $q$ refers to the recommendations.

Therefore popularity lift (PL) for group $G$ is defined as:
\begin{equation}
    PL(G)=\frac{GAP_q(G)-GAP_p(G)}{GAP_p(G)}
\end{equation}

Positive values for $PL$ indicate amplification of popularity bias by the algorithm. A negative value for $PL$ happens when, on average, the recommendations are less concentrated on popular items than the users' profile. Moreover, the $PL$ value of 0 means there is no popularity bias amplification.

Our goal is to show how popularity lift, $PL$, and miscalibration, $MC$ are correlated for different algorithms. In particular, we are interested in discovering how popularity bias in recommendation algorithms affects groups of users with different degrees of interest in popular items. That will reveal if there is any discrimination in terms of imposing popular items on different user groups. In addition, we want to show how the degree of being affected by algorithmic popularity bias (i.e. popularity lift) for these groups is correlated with how much miscalibration they experience in their recommendations.

\section{Results}

\subsection{Popularity Bias Propagation}
\begin{figure*}
\centering
\SetFigLayout{5}{1}
  \subfigure[BMF]{\includegraphics[width=1.9in]{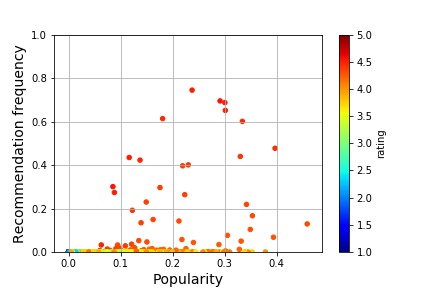}}
\subfigure[SVD++]{\includegraphics[width=1.9in]{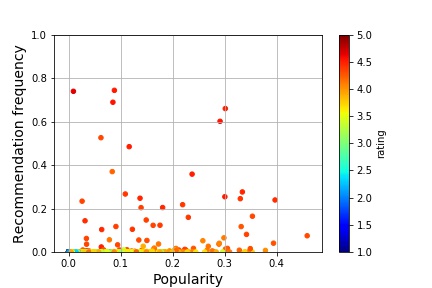}}
\subfigure[ItemKNN]{\includegraphics[width=1.9in]{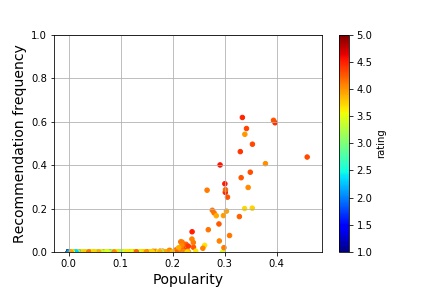}}
\subfigure[UserKNN]{\includegraphics[width=1.9in]{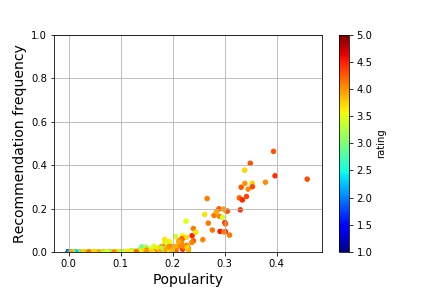}}
  \subfigure[Most-popular]{\includegraphics[width=1.9in]{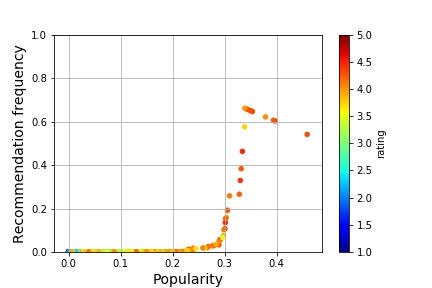}}

\captionsetup{justification=centering}
\caption{MovieLens:\\The correlation between the popularity of items and how frequent they are being recommended.}
\label{movielens_corr}
\end{figure*}

\begin{figure*}
\centering
\SetFigLayout{5}{1}
  \subfigure[BMF]{\includegraphics[width=1.9in]{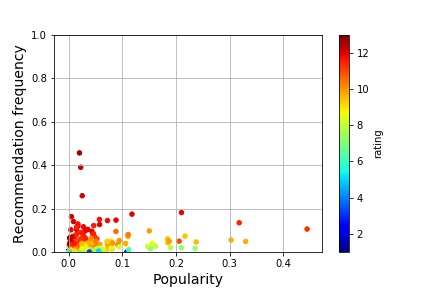}}
  \subfigure[SVD++]{\includegraphics[width=1.9in]{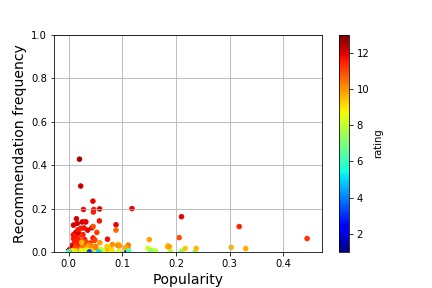}}
 \subfigure[ItemKNN]{\includegraphics[width=1.9in]{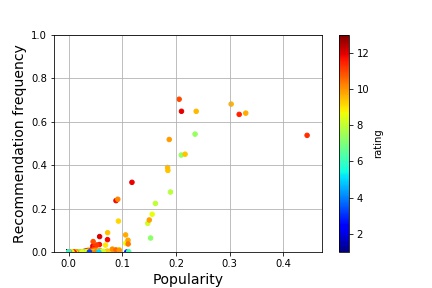}}
 \subfigure[UserKNN]{\includegraphics[width=1.9in]{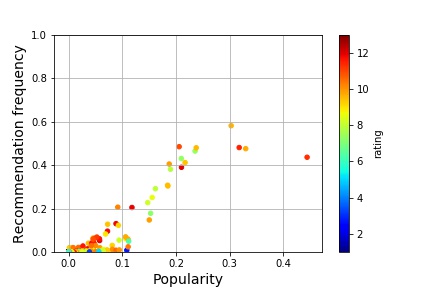}}
\subfigure[Most popular]{\includegraphics[width=1.9in]{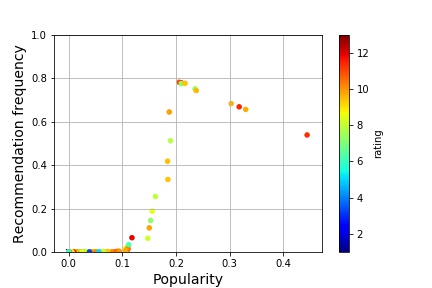}}

\captionsetup{justification=centering}
\caption{Yahoo Movies:\\The correlation between the popularity of items and how frequent they are being recommended.}\label{yahoo_corr}
\end{figure*}

Not every recommendation algorithm has the same amplification power for popularity bias. We first look at the performance of different algorithms in terms of popularity bias propagation without paying attention to how they perform for different users or groups of users. 

Figure ~\ref{movielens_corr} shows the correlation between the number of times an item is rated and how frequent it is recommended by different algorithms on MovieLens dataset. It can be seen that in all algorithms, there are many items that are almost never recommended (the items fall on the horizontal tail of the scatter plot). As expected the \textit{Most popular} algorithm seems to have the strongest correlation between the number of times an item is being rated and the number of times it is recommended. Both \textit{User KNN} and \textit{Item KNN} also show strong correlations. 

Looking at the plot for the \textit{BMF} algorithm, it seems there is not a strong correlation between how often an item is rated and how often it is recommended. Nevertheless, its very sparse scatter plot shows the number of recommended items is still low. In this algorithm, there are many items that are recommended very rarely (even some of those that are rated more frequently) while few items are being recommended frequently. The same fact applies to \textit{SVD++}. For each point on the scatter plot, you can also see the average rating for the corresponding item to illustrate the quality of these recommended items using each algorithm. 

Figure ~\ref{yahoo_corr} shows the same information on Yahoo Movies data. Note that, for all algorithms, items that are already rated by a user are not recommended and that is why there is an inflection point around the popularity value of 0.35 for MovieLens and 0.2 for Yahoo Movies on the plot for the \textit{Most popular} algorithm. 

\begin{figure}
\centering
\SetFigLayout{1}{2}
  \subfigure[MovieLens]{\includegraphics[width=1.7in]{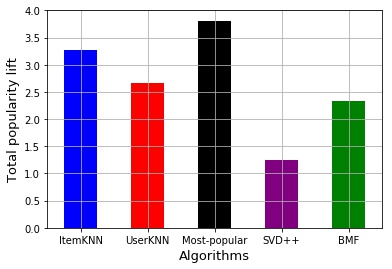}}
  \hfill
  \subfigure[Yahoo Movies]{\includegraphics[width=1.7in]{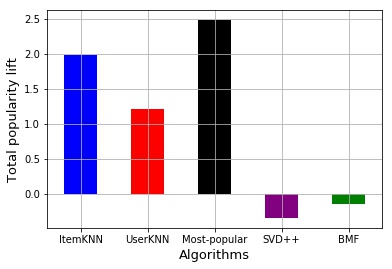}}
  \hfill
\caption{Total popularity lift for different recommendation algorithms.}
\label{total_poplift}
\end{figure}

Figure ~\ref{total_poplift} shows the total popularity lift for different algorithms on MovieLens and Yahoo Movies datasets. As can be observed, the most popular algorithm has the highest popularity lift followed by $ItemKNN$ and $UserKNN$ on both datasets. \textit{SVD++} and $BMF$ show a negative popularity lift on the Yahoo Movies dataset meaning the recommendations, on average, are less concentrated on popular items for different user groups compared to their profiles. That is because these two algorithms do not have much popularity bias on Yahoo Movies data and are generally recommending items with smaller average popularity to users contributing an overall negative popularity lift.  However these two algorithms still show positive popularity lift on MovieLens dataset, though to a much lesser degree than their neighborhood-based counterparts. 

\begin{figure}
\centering
\SetFigLayout{1}{2}
  \subfigure[MovieLens]{\includegraphics[width=2in]{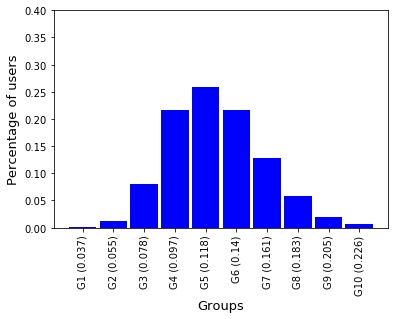}}
  \hfill
  \subfigure[Yahoo Movies]{\includegraphics[width=2in]{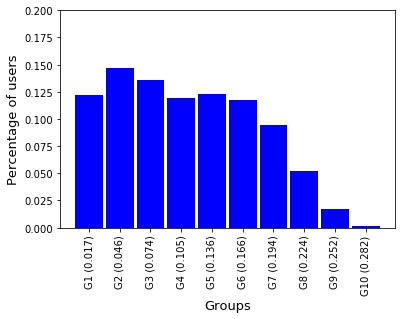}}
  \hfill
\caption{Histogram of group's average item popularity in MovieLens and Yahoo Movies datasets.}
\label{group_propensity_hist}
\end{figure}

Figure ~\ref{group_propensity_hist} shows the impact of popularity bias on the groups we created based on their degree of interest towards popular items. The average popularity of each group is shown within the parentheses next to each group's label with smaller values representing groups with more niche tastes and larger values for more blockbuster-focused groups. The percentage of users that fall within each group is shown on the y-axis of this figure. We can see that for MovieLens dataset, there is a normal distribution which reaches its peak between 0.10 and 0.13 while there are a few users who fall within the more extreme groups on the left and right side of the plot. For Yahoo Movies dataset, however, the percentage of users on each group is more consistent except for the groups with higher average popularity where there are only a few users. 

\subsection{Relationship Between Popularity Bias and Miscalibration}

In this part we show the connection between popularity bias and unfairness in recommendation. Making this connection would be helpful in many fairness-aware recommendation scenarios because fixing the popularity bias could be used as an approach to tackle this type of unfairness.

\begin{figure}
\centering
\SetFigLayout{1}{2}
  \subfigure[MovieLens]{\includegraphics[width=2.5in]{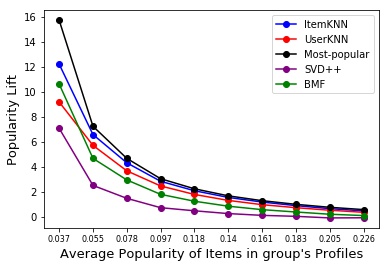}}
  \hfill
  \subfigure[Yahoo Movies]{\includegraphics[width=2.5in]{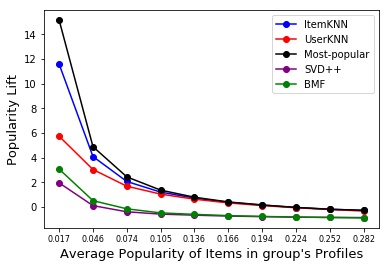}}
  \hfill
\caption{Average item popularity of user groups (G1 through G10 from left to right) and their observed popularity lift in MovieLens and Yahoo Movies datasets. }
\label{pop_lift_pop}
\end{figure}

An illustration of the effect of the algorithmic popularity bias on different user groups is shown in figure ~\ref{pop_lift_pop}. Each dot represents a group with certain average popularity of the users' profiles in that group which is shown on the x-axis. On the y-axis, the popularity lift of different algorithms on each user group is depicted. It can be seen that groups with the lowest average popularity (niche tastes) are being affected the most by the algorithmic popularity bias and the higher the average popularity of the group, the lesser the group is affected by the popularity bias. This shows how, unfairly, popularity bias is affecting different user groups. Table ~\ref{table_pop_lift_extreme} shows the popularity lift experienced by two extreme groups: $G_1$ (the most niche group) and $G_{10}$ (the most blockbuster-focused group). It can be seen that, for all of the algorithms, the difference in popularity lift experienced by $G_1$ is significantly higher than $G_{10}$ ($p<0.01$ for T-test significance). 

\begin{figure}
\centering
\SetFigLayout{1}{2}
  \subfigure[MovieLens]{\includegraphics[width=2.4in]{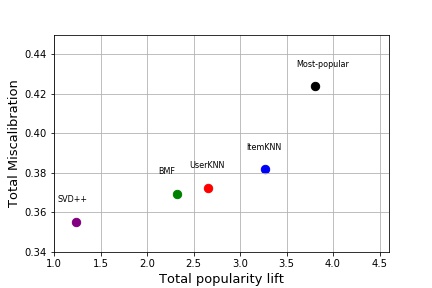}}
  \hfill
  \subfigure[Yahoo Movies]{\includegraphics[width=2.4in]{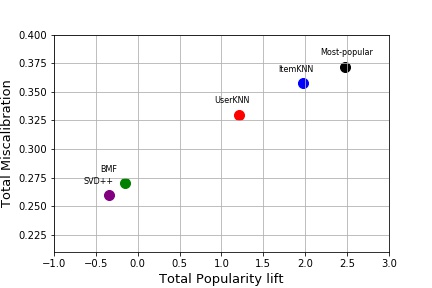}}
  \hfill
\caption{Comparison of different algorithms in terms of their total popularity lift and miscalibration.}
\label{pop_miscal}
\end{figure}

Figure \ref{pop_miscal} shows the correlation between total popularity lift and miscalibration for different algorithms on MovieLens and Yahoo Movies datasets. We can see a general trend for these algorithms: The more an algorithm imposes popularity bias (highest total popularity lift) on users, the higher the degree of miscalibration, showing a positive correlation between these two. Algorithms with the highest popularity lift such as $UserKNN$ and $ItemKNN$ also have a higher miscalibration compared to other algorithms. \textit{SVD++} and  $BMF$ have the lowest popularity lift and they also have the lowest miscalibration. This is an interesting finding as it could be useful in the design of recommender systems where sensitivity to this type of miscalibration may dictate the choice on the underlying algorithm used.

Table ~\ref{table_miscalibration_extreme} shows the miscalibration values for two groups $G_1$ and $G_{10}$ (two groups on the extreme sides of the popularity spectrum, as we discussed earlier). It can be seen that, for all algorithms, the miscalibration value for $G_1$ is significantly higher than the one for $G_{10}$ ($p<0.01$ for T-test significance). This shows the group with the lowest value for its average item popularity has experienced the highest miscalibration for their recommendations. Moreover, we also saw in Table ~\ref{table_pop_lift_extreme} that this group experienced the highest popularity lift as well. This shows again how popularity lift might lead to miscalibration.

\begin{table*}[t]
    \centering
    \caption{\label{tab:results} The popularity lift of different recommendation algorithms on two groups $G_1$ and $G_{10}$}
    \begin{tabular}{ l c c c c }
        \toprule
        & \multicolumn{2}{c}{MovieLens}  & \multicolumn{2}{c}{Yahoo Movies} \\
        \cmidrule(lr){2-3}\cmidrule(lr){4-5}
            & \thead{Popularity lift for \\ $G_{10}$} & \thead{Popularity lift for \\ $G_1$} & \thead{Popularity lift for \\  $G_{10}$} & \thead{Popularity lift for \\  $G_1$}  \\
            \midrule
        ItemKNN  &    
                0.4579     &  12.19         &  -0.26 &  11.57  \\
                    \bottomrule
        UserKNN      & 0.348 &  9.17    & -0.31 & 5.738  \\
          \bottomrule
       Most-popular &      
       0.563  & 15.7   &  -0.25 & 15.13 \\
    \bottomrule
     SVD++   
     &  -0.09 &  7.063  &  -0.84 &1.96 \\
    \bottomrule
   BMF  
    &  0.086  &  10.60  &  -0.871 &3.079 \\
    \bottomrule
    \end{tabular}
\label{table_pop_lift_extreme}
\end{table*}

\begin{table*}[t]
    \centering
    \caption{\label{tab:results} The miscalibration of different recommendation algorithms on two groups $G_1$ and $G_{10}$}
    \begin{tabular}{ l c c c c }
        \toprule
        & \multicolumn{2}{c}{MovieLens}  & \multicolumn{2}{c}{Yahoo Movies} \\
        \cmidrule(lr){2-3}\cmidrule(lr){4-5}
            & \thead{Miscalibration for \\ $G_{10}$} & \thead{Miscalibration for \\ $G_1$} & \thead{Miscalibration for \\  $G_{10}$} & \thead{Miscalibration for \\  $G_1$}  \\
            \midrule
        ItemKNN  &    
                0.250    &  0.418        &  0.345 &  0.4660  \\
                    \bottomrule
        UserKNN      &0.248&  0.446   & 0.345 & 0.3953 \\
          \bottomrule
       Most-popular &      
       0.277  & 0.501   & 0.342 & 0.471\\
    \bottomrule
     SVD++     
     &  0.380 &  0.556  &  0.272 &0.315 \\
    \bottomrule

   BMF  
    & 0.396  &  0.635  & 0.290 &0.345 \\
    \bottomrule
    \end{tabular}
    \label{table_miscalibration_extreme}
\end{table*}

\section{Gender Analysis}
The grouping methodology based on users' interest in popular items was to illustrate how different levels of interest in item popularity affects the recommendations and miscalibration. In particular, we intended to show how different recommendation algorithms are treating different user groups with varying degree of interest in popular items.

In this section we use gender as the grouping criteria to analyse the impact of popularity bias on different user groups and its correlation with the miscalibration of the recommendations for these two groups. Between the two datasets we used in previous sections only MovieLens has gender information about the users. Therefore, in this section, we only reported the results for different algorithms using this dataset. Out of 6040 users in this dataset, 1708 are women and 4330 are men which shows an imbalance nature of this dataset. Table ~\ref{table_pop_lift_gender} shows the popularity lift of different recommendation algorithms for men and women. It can be observed that for all algorithms, women have experienced a significantly higher degree of popularity lift ($p<0.05$ except for SVD++ which treated both groups equally). What is interesting and is also consistent with our results in Section 5 is that the group who experienced higher popularity lift (women) also experienced higher miscalibration, as can be seen in Table ~\ref{table_miscalibration_gender}. The popularity lift and miscalibration of the SVD++ for men and women are very close to each other which shows the consistency of this algorithm in delivering fairer recommendations. BMF imposes higher popularity lift on women but it treats both men and women equally when it comes to miscalibration. Note that, the larger group size for men (4330 vs 1708) could be another reason for higher miscalibration for women as the algorithms are trained on a large number of ratings for men and that could make the algorithms biased towards learning men's rating behavior. Further analysis regarding the rating behavior of these two groups such as consistency of their ratings and the informativeness of their profile would help in shedding light on the results reported here. 

\begin{table*}[t]
    \centering
    \caption{\label{tab:results} The popularity lift of different recommendation algorithms on two groups men and women.\\ Bold values show significance difference with $p<0.05$}
    \begin{tabular}{ l c c c}
        \toprule
        & \multicolumn{2}{c}{MovieLens} \\
        \cmidrule(lr){2-3}
            & \thead{Popularity lift for \\ men} & \thead{Popularity lift for \\ women}  \\
            \midrule
        ItemKNN  &    
                \textbf{1.58}     &  \textbf{1.76}  \\
                    \bottomrule
        UserKNN      & \textbf{1.33} &  \textbf{1.52} \\
          \bottomrule
       Most-popular &      
       \textbf{1.71}  & \textbf{1.91}  \\
    \bottomrule
     SVD++   
     &  0.33 &  0.33  \\
    \bottomrule
   BMF  
    &  \textbf{0.87}  &  \textbf{1.01}  &  \\
    \bottomrule
    \end{tabular}
\label{table_pop_lift_gender}
\end{table*}
\begin{table*}[t]
    \centering
    \caption{\label{tab:results} The miscalibration of different recommendation algorithms on two groups men and women. \\Bold values show significance difference with $p<0.05$}
    \begin{tabular}{ l c c c}
        \toprule
        & \multicolumn{2}{c}{MovieLens} \\
        \cmidrule(lr){2-3}
            & \thead{Miscalibration for \\ men} & \thead{Miscalibration for \\ women}  \\
            \midrule
        ItemKNN  &    
                \textbf{0.368}    &  \textbf{0.42} \\
                    \bottomrule
        UserKNN      &\textbf{0.36}&  \textbf{0.4}  \\
          \bottomrule
       Most-popular &      
       \textbf{0.40} & \textbf{0.48} \\
    \bottomrule
     SVD++     
     &  0.41&  0.43  \\
    \bottomrule

   BMF  
    & 0.37  &  0.37 \\
    \bottomrule
    \end{tabular}
    \label{table_miscalibration_gender}
\end{table*}

\section{Discussion}
In Figure 9 we observed that there is a positive correlation between popularity lift and the overall miscalibration of a recommendation algorithm. Investigating which one is causing the other one needs a more in-depth experiment design and further analysis. However, we believe that popularity lift causes miscalibration and not the other way around. When a list is miscalibrated, it is due to over-representation or under-representation of some genres relative to what the user expects given the items that s/he has rated. Popularity lift increases the recommendation frequency of popular movies and the genres associated with them. As a result, these popular genres become over-represented (at the cost of suppressing the non-popular movies and their associated genres) and thus cause overall miscalibration. 


We also observed in Table 1 that user groups who experience higher popularity lift also experience higher miscalibration. As we saw in Figure 8, user groups with lesser interest in popular items are affected more by the popularity bias of a recommendation algorithm (i.e. higher degree of popularity lift), because their interests are less likely to lie within the set of popular items. The mechanism suggested above would imply that these groups would also experience greater miscalibration, and that is in fact what we found.


\section{Conclusion and Future Work}

Recommendation algorithms often suffer from popularity bias problem. This bias could be problematic for reasons such as the need for generating a fuller understanding of users' preferences which is usually done by recommending non-popular items and, to avoid ignoring certain items while giving too much attention to few items which is referred to as social good. 

In this paper, we looked at the popularity bias problem from the user's perspective and we observed different groups of users can be affected differently by this bias depending on how much they are interested in popular items. We also showed that the popularity bias has a strong correlation with miscalibration which measures how consistent the recommendations are with the true users' preferences. That is, algorithms with high popularity bias (popularity lift) tend to also have higher total miscalibration. In particular, for two extreme groups on the spectrum of item popularity, we showed that the group which is less interested in popular items is affected the most by popularity bias and also has the highest level of miscalibration. Consistent with these results, we observed that on MovieLens dataset the group who experienced higher popularity lift (women) also experienced higher miscalibration compared to men. 

We also showed that different algorithms behave differently with respect to miscalibration. In particular, commonly used neighborhood-based algorithms are highly susceptible to propagating popularity bias and hence to miscalibration. On the other hand, factorization based algorithms seem more resistant to this effect. 

One limitation for our work in this paper is that all of our experiments only indicated a correlation between popularity bias and miscalibration and unfairness. For future work, we intend to study the causality of popularity bias on these issues. In particular, we will design experiments such as sampling methods to control popularity bias in data and see the effect of that on miscalibration and fairness. We will also investigate the effect of algorithms for controlling algorithmic popularity bias on miscalibration and fairness. 

\bibliographystyle{ACM-Reference-Format}
\bibliography{ref.bib}

\end{document}